\documentstyle[prd,aps]{revtex}
\begin{document}
\draft
 \twocolumn[\hsize\textwidth\columnwidth\hsize\csname
 @twocolumnfalse\endcsname
\preprint{SU-ITP-97/46, hep-th/9801073,
January 1998}

\title { \bf  Pre-Big-Bang  Requires the Universe to be Exponentially
Large From the Very Beginning}

\author{Nemanja Kaloper, Andrei Linde, and Raphael Bousso}
\address{Department of Physics, Stanford University, Stanford, CA
94305-4060, USA}
\date {\today}
\maketitle
\begin{abstract}
We show that in a generic case of the pre-big-bang scenario, inflation will
solve cosmological problems only if the universe at the onset of inflation is
extremely large and homogeneous from the very beginning. The size of a
homogeneous part of the universe at the beginning of the stage of pre-big-bang
(PBB) inflation must be greater than $10^{19}$ $l_s$, where $l_s$ is the
stringy length, which is the only natural length scale in PBB cosmology. The
total mass of an inflationary domain must be greater than $10^{72} M_{s}$,
where $M_{ s} \sim l_s^{-1}$. If the universe is initially radiation dominated,
then its total entropy at that time must be greater than $10^{68}$. If the
universe is closed, then at the moment of its formation it must be uniform over
$10^{24}$ causally disconnected domains. The natural duration of the PBB stage
in this scenario is $M_p^{-1}$.  If the universe is open, then its initial
state should have been very homogeneous over an infinitely large distance, in
order to account for the homogeneity of our  part of the universe.   We argue
that  the initial state of the open PBB universe could not be homogeneous
because of quantum fluctuations.
Independently of the issue of homogeneity,
one must introduce two large dimensionless parameters, $g_0^{-2} > 10^{53}$,
and $B > 10^{38} g_0^{-2} > 10^{91}$, in order  to solve the flatness problem
in the PBB cosmology.  A regime of eternal inflation does not occur in the PBB
scenario.  This should be compared with the simplest versions of the chaotic
inflation scenario, where the regime of eternal inflation may begin in a
universe of size $O(M_{ p}^{-1})$ with vanishing initial radiation entropy,
mass $O(M_p)$, and geometric entropy $O(1)$. We conclude that the current
version of the PBB scenario cannot replace usual inflation even if one solves
the graceful exit problem in this scenario.
\end{abstract}

\

 \vskip2pc]
\section{Introduction}

After 15 years of development of the theory of inflationary universe,
cosmologists believe that something like inflation is indeed necessary
in order to construct an internally consistent cosmological theory.
The best hopes for the theory of all fundamental interactions are
related to superstring theory.  Unfortunately it has been known for at
least a decade that it is extremely hard to construct a working
mechanism for inflation in string theory. The familiar inflationary
scenarios, which have been developed in the context of general
relativity or even its scalar-tensor extensions, cannot be derived
from the string effective action on a generic background. The reason
is that in the low energy effective action the string moduli, and in
particular the dilaton field, couple nonminimally to other degrees of
freedom, and hence run during cosmological evolution. The running of
the dilaton slows down the expansion of the horizon, and as a result,
the horizon and flatness problems cannot be solved.

Hopefully, this problem is only temporary; it may be related to our
limited understanding of string theory, which rapidly changes every
year. It might be possible also, that inflation can be implemented in
the simplest versions of string theory, but in a rather nontrivial
way. One of the most interesting suggestions in this respect is the
pre-big-bang (PBB) scenario developed in \cite{prebb}.  This scenario
assumes that inflation occurred in a stringy phase prior to the big
bang. It relies on the running of the dilaton, which at some time
starts to dominate the evolution, and causes the scale factor to
diverge in the future. Further, the scenario assumes that there exists
a mechanism which eventually becomes important and overturns the
dilaton-dominated expansion into a simple power-law, thus avoiding the
big bang singularity and solving the horizon problem at the same time.
The exit, however, is difficult to attain. For example, it is known,
thanks to several exact no-go theorems \cite{bv,kmo1,rest}, that it
cannot be obtained in the effective potential approximation. As a
result, the current version of PBB holds that the exit should occur in
the strong coupling/large curvature sector of string theory, where
higher order quantum corrections become important.

The PBB scenario has some other problems. It is not easy to obtain
density perturbations with a nearly-flat spectrum, it is not quite
clear how to solve the primordial monopole problem, etc.  Despite
these difficulties, the possibility that PBB cosmology may provide a
realization of inflationary cosmology in the context of some
string-inspired models is certainly very interesting. However,
recently it has been argued by Turner and Weinberg \cite{tw} that
there also exists a fine-tuning problem in the PBB scenario. If the
dilaton-dominated era in PBB is preceded by a non-inflating phase,
then in order for inflation to solve the horizon and flatness
problems, the initial universe at the onset of the dilaton-dominated
era should be very large. The authors of Ref.~\cite{tw} concluded that
PBB is less robust, and therefore less attractive as an implementation
of the inflationary paradigm.  On the other hand, from Ref.~\cite{tw}
it was not quite obvious how strong this fine-tuning should be, and
whether one can avoid this problem altogether if the universe is open
\cite{ven,buon}, or if one somewhat modifies the PBB scenario
\cite{ms}. In fact, the largeness of the size of the PBB universe was
noticed in the very first papers on the PBB model \cite{prebb}, but it
was not considered a real problem.

More recently, another attempt has been made \cite{coule} to argue
that PBB is unnatural essentially because at the time of the exit the
universe should be very large. However, in our opinion, the author of
Ref.~\cite{coule} somewhat misinterpreted the PBB theory and ignored
the possibility that this problem can be solved by PBB inflation.

In this paper we will discuss the fine-tuning problem in PBB, using
only the most general and widely accepted premises: (1) generic
initial conditions which allow that the universe has not been
dilaton-dominated throughout its past, (2) the existence of an
inflationary phase, which metamorphoses into a post-inflationary
universe like our own via a successful process of branch-changing, and
(3) the validity of the description of the model by string theory with
higher order corrections, implying that $g_s \leq1$. Using this, we
will show that if our part of the universe appeared as a result of
pre-big-bang inflation, then it should have originated from a
homogeneous domain of exponentially large initial size $L_{i} >
10^{19} l_s$, where $l_s = M^{-1}_s$ is the stringy scale. Such a
domain should have enormously large initial mass and initial entropy
at the onset of PBB inflation.

Immediate consequences of this result look different for a closed
universe and for an open one, but the main conclusions are very
similar. A closed universe appears from a singularity, and its
description in terms of the effective action used in the PBB theory
becomes possible only at a stringy time $t_s = l_s = M^{-1}_s$ after
the singularity. We will show that at that time the PBB universe must
consist of at least $10^{24}$ causally disconnected regions with
nearly equal density. The emergence of such a huge homogeneous
universe is extremely improbable; this is the well-known horizon
problem which inflationary theory is supposed to solve. To solve this
problem in the context of the PBB cosmology one would need an
additional stage of inflation, not related to the PBB scenario.

If the PBB universe is spatially open, it does not have an initial
singularity.  Instead, it starts as an infinitely large patch of
Minkowski space (or, more precisely, the Milne universe) in the very
far past, with almost vanishing matter content except for an
infinitesimally small, spatially homogeneous dilaton kinetic energy
density.  Such a universe should shrink for an infinitely long time,
until the dilaton density grows sufficiently large to cause the scale
factor to bounce and undergo superinflation.

The initial homogeneity of an open PBB universe filled by a dilaton
field $\Phi$ with an infinitesimally small energy density can be
completely destroyed even by extremely small but finite classical
perturbations of the dilaton field.  Thus, in order to explain why our
universe is large, one would need to assume that it is infinite, and
in order to explain homogeneity of our local part of the universe, one
would need to assume that the early universe was relatively
homogeneous on an infinitely large scale. Moreover, we will argue that
the initial density of the open universe was so small that its
homogeneity could be easily destroyed by quantum fluctuations which
were present in the very early universe.

Of course, if the universe is infinite, one can always find a
sufficiently large and homogeneous patch of spacetime which can
undergo PBB inflation \cite{buon}. Parts of the universe similar to
the one where we live now can appear as a result of PBB inflation of
homogeneous domains with an initial size greater than $10^{19} \ l_s$.
However, one may argue that in a contracting universe with chaotic
initial conditions, the only natural size of a homogeneous region is
$l_s$.  Therefore, the spontaneous formation of a homogeneous domain
of size $10^{19}$ $ l_s$ seems extremely improbable.

We will show also that  even if it were possible to solve the homogeneity
problem in an open PBB universe,   the possibility to solve the flatness
problem  relies on the existence of two large dimensionless parameters,
$g_0^{-2} > 10^{53}$, and $B > 10^{38} g_0^{-2} > 10^{91}$. Thus, in order to
explain why our universe is flat, one must first explain  the origin of these
two large dimensionless parameters.

A distinguishing feature of many inflationary models is the existence
of a regime of self-reproduction of inflationary domains. This regime
leads to especially profound consequences in the chaotic inflation
scenario \cite{al,book}, but it occurs in many other inflationary
models as well \cite{Vilen}.  The self-reproduction of inflationary
domains may alleviate the initial condition problem even for those
inflationary models where the initial conditions required for
inflation are unnatural \cite{LLM}.  We have investigated the
possibility that a similar mechanism might alleviate the problem of
initial conditions in the PBB inflationary cosmology, but we have
found that the dynamics of PBB precludes the possibility of
self-reproduction.

\section{Pre-Big-Bang}

We begin here by reviewing the PBB scenario, to the extent needed for
our investigation. The dynamics of the model is given by the
four-dimensional effective action of string theory, which contains the
metric, dilaton and possibly other matter fields. In the string
frame, to the lowest order, the action can be written as \cite{prebb}
\begin{equation}
\label{act}
S = \int d^4 x \sqrt{|g_{\mu\nu}|}\
 \frac{e^{-\sigma}}{  l_s^2}\,\left[{1\over 2} R +
{1\over 2}(\nabla
\sigma)^2 + {\cal L}_m({\cal Y}, g_{\mu\nu}, \sigma) \right] .
\end{equation}
where $\sigma$ is the dimensionless dilaton field, which determines
the string coupling constant $g = \exp(\sigma/2)$, $g_{\mu\nu}$ is the
string-frame metric and ${\cal Y}$ denotes any additional matter
degrees of freedom. The dimensional parameter $l_s$ is the string
scale, given by $l_s \sim \sqrt{\alpha'}$, and is close to today's
value of the Planck scale. We set $\hbar = c =1$, so that mass, time
and length all have the same units, given by $l_s$. The simplest
variant of PBB has been given for spatially flat FRW backgrounds, and
has since been investigated in more complicated situations. The crux
of the scenario is that any solution starting in the weak coupling
regime $g_s \ll1$ always evolves such that the dilaton grows (i.e.,
the coupling increases), until it dominates the evolution. Then the
universe begins the accelerated expansion, since the scale factor
grows superexponentially towards a singularity in the future. The
scenario further needs a mechanism, dubbed branch-changing, which can
overturn the dilaton-dominated expansion into a simple power-law, thus
avoiding the singularity and solving the horizon problem at the same
time. While this cannot be attained in the effective potential limit
of string theory, where the inflating solutions always run to the
future curvature singularity \cite{bv,kmo1,rest}, it has been
suggested that string-scale physics could yield the exit
\cite{venetal,bm}. Since none of our arguments require details of the
exit scenario, we will not discuss the prospects for actually
constructing it. We will, however, underline the basic assumptions
made in positing that an exit is possible.

Let us now look at the salient features of the dilaton-dominated phase
in PBB cosmology. We can easily write the equations of motion which
follow from the action (\ref{act}). In the presence of matter sources,
they are\footnote{Our normalizations are such that the factor $16\pi$
  which occurs in \cite{tw} has been absorbed into $l_s^2$.}
\cite{prebb}
\begin{eqnarray}     \label{ee1}
&&\ddot \sigma + 3H \dot \sigma - {\dot \sigma}^2 =
l^2_s e^{\sigma}(3p - \rho) \nonumber \\
&&\dot H = H \dot \sigma - 3 H^2 - 2\frac{k}{a^2}+ l_s^2
p e^{\sigma}
\nonumber \\
&& \dot \rho + 3 H (\rho + p) = 0 \\ && {\dot \sigma}^2 + 6 H^2 - 6
H \dot \sigma + 6 \frac{k}{a^2}- 2 l^2_s e^{\sigma} \rho = 0
\nonumber
\end{eqnarray}
where $H=\dot a/a$ is the string-frame Hubble parameter, and $\rho$
and $p$ are the effective matter density and pressure. For most
practical purposes, it is sufficient to assume a simple equation of
state for the matter, $p = \gamma \rho$ with $\gamma \in [-1/3,1/3]$
for classical stringy matter. We will not delve into the realm of
exotic degrees of freedom which might arise at very large curvature or
strong coupling, and which might require more complicated equations of
state. This will not restrict the generality of our arguments, since
we will be studying the conditions at or before the onset of
inflationary expansion. There, the validity of the effective action
implies that matter behaves ordinarily.

After the onset of the inflationary phase, where all other
contributions (matter, spatial curvature etc.) to (\ref{ee1}) except
the dilaton are negligible, one can easily find the pure
metric-dilaton solutions \cite{prebb}. The expanding solutions are
divided into two classes separated by the curvature singularity. For
$t < 0$, one has
\begin{equation}
\label{sol1}
 a_+ = a_0 (-t)^{-\frac{1}{\sqrt{3}}},
 ~ H_+ = \frac{1}{\sqrt{3}(-t)}, ~
e^{-\sigma_+} = e^{-\sigma_0} (-t)^{1+\sqrt{3}},
\end{equation}
 whereas for $t > 0$, one has
\begin{equation}\label{sol2}
 a_- = a_0 t^{\frac{1}{\sqrt{3}}}, ~~ H_- = \frac{1}{\sqrt{3}t}, ~~~
e^{-\sigma_-} = e^{-\sigma_0} t^{1-\sqrt{3}}.
\end{equation}
The solutions for $t<0$ are by now widely referred as to the $(+)$, or
superinflationary, branch, and those for $t>0$ as the $(-)$, or
post-inflationary, branch\footnote{The proper definition of branches
  is derived from solving the quadratic constraint equation in
  (\ref{ee1}) for $\dot \sigma$, and the sign of each branch is
  determined by the sign of the square root which arises in the
  solution.}. For later purposes, it will be useful to re-express the
$(+)$-branch solution in terms of the effective Planck mass squared,
\begin{equation}
\label{plm}
\Phi = M^2_p(t) =  \frac{1}{l^2_{p}(t)} = {l_s^{-2}} e^{-\sigma} \ ,
\end{equation}
as follows:
\begin{equation}
\label{mc+}
a_+ = |t_0| \left|\frac{t_0}{t}\right|^{1/\sqrt{3}}, ~~~~~~~
\Phi = \frac{1}{l^2_{p}(0)} \left|\frac{t}{t_0}\right|^{\sqrt{3}+1},
\end{equation}
where the initial time $t_0$ is the aforementioned time of the onset
of the pole-dominated era.

Note that the hypersurface $t=0$ is singular, lying in the future of
the $(+)$ (inverse power-law) branch solutions and in the past of the
$(-)$ (power-law) branch solutions. The role of the branch-changing
mechanism is to connect the two branches, $(+)$ chronologically
preceding the $(-)$, in such a way as to remove the singularity, and
to allow the superexponential growth of the $(+)$ branch to play the
role of inflation. For simplicity, in this article we will assume that
such a matching of solutions is possible, without delving into a
detailed analysis of string physics. However, we need to review
carefully the phenomenological aspects of branch-changing. The current
lore of PBB is to assume that the exit must occur roughly when the
description given by the effective action (\ref{act}) breaks down, but
that the dynamics can be described by a sum of (\ref{act}) and higher
order corrections, both in $g_s$ and in $\alpha'$. Since the
lowest-order string effective action (\ref{act}) is a truncation of a
double expansion in both the inverse string tension $\alpha'$ and the
string coupling $g_s = \exp(\sigma/2)$, the description based on it
breaks down either when $R \sim 1/\alpha'$ or when $g_s \sim 1$,
whichever happens first. But the higher genus terms alone cannot
saturate the growth of the curvature \cite{kmo1,venetal,bm}, and so
even if the exit is precipitated by $g_s \sim 1$, the curvature would
continue to grow until again $R \sim 1/\alpha'$. Similarly, the higher
derivative terms cannot complete the exit by themselves, and thus if
the breakdown of the action (\ref{act}) is caused by $R\sim 1/\alpha'$
effects, the exit does not occur until the coupling reaches $g_s \sim
1$ \cite{venetal,bm}, while the curvature remains approximately
constant. If this happens, the dilaton-driven inflation is supplanted
at a late time by an exponential phase \cite{venetal,bm,ms}. In any
case, all exit scenarios in PBB postulate that at the time of the exit
the Hubble length must be of the order of string scale, $H^{-1} \sim
l_s$, while the string coupling is always assumed to be at most of
order unity\footnote{One can consider $\Phi/H^2 \sim (1/l^2_{p}(0))
  |t|^{3+\sqrt{3}}/|t_0|^{\sqrt{3}+1}$. Initially, $\Phi_i/H^2_i \sim
  |t_i|^2/l^2_{p}(t_i) \gg1$. Further, $\Phi/H^2$ is a monotonically
  decreasing function during the dilaton-dominated era, and becomes of
  order unity at a time $t_{*}\sim t_0^{1/\sqrt{3}}
  l^{1-1/\sqrt{3}}_{p}(0)$. This time must be chronologically after
  the string scale $l_s$. Indeed, suppose the opposite, that $t_*$
  precedes $l_s$. Since at the exit roughly $H_f \sim 1/l_s$, it would
  then be $\Phi_f \ll H_f^2 \sim l_s^{-2}$, or since $g_s(t_f) =
  \exp(\sigma_f/2) = l_s/\sqrt{\Phi_f}$, the coupling would be
  $g_s(t_f) \gg1$, contradicting the assumption that the effective
  action is well defined. Hence, $t_{*}$ must be after the string
  scale $l_s$, and so, at the moment of the exit, it must be $\Phi_f
  \le l_s^{-2}$.}. Therefore, using the subscripts $i$ and $f$ to
denote the beginning and the end of the inflationary era, we see that
when the exit occurs it must be
\begin{equation}
\label{exit}
H_f^{-1} \sim |t_f| \sim l_s ~,~~~~~~~ \Phi_f \le {l^{-2}_s} \ .
\end{equation}
This illustrates how the conclusion in fact hinges on the assumption
that $g_s \le 1$. Given this, the numerical values of any bound which
we will obtain using (\ref{exit}) will be very robust.

After the exit has been completed, the solution continues as a $(-)$
branch. Hence, from our vantage point, the Big Bang would really
correspond to the emergence of the universe from the string phase, at
the time roughly of order of $l_p$. Hereafter, we can for simplicity
assume ``maximal decoupling" of the dilaton, taking it to be constant
from the Planck time onwards. This is phenomenologically desirable,
since at present the Planck scale is roughly the same as the string
scale, $l_p \sim l_s$, while the string coupling is of order unity,
and after the Big Bang neither of these quantities could have changed
very much, if baryogenesis and nucleosynthesis were to proceed
normally \cite{co}. In order to determine if PBB is a viable
inflationary model, we need to check if the dilaton-dominated
expansion solves cosmological problems. The resolution of the horizon
problem requires that inflation naturally produce the ``initial"
condition that the size of the observable part of the universe
extrapolated back to the Planck time is greater than the corresponding
Planck length by about $30$ orders of magnitude: $L_f\ge 10^{30} l_p
\sim e^{70} l_p$. The initial size of the homogeneous part of the
universe $L_i$ at the beginning of PBB inflation should be greater
than the Hubble length $H^{-1}_i \sim |t_i|$, because only in this
case inflation ensures homogeneity of the part of the universe
produced by the expansion of the original domain.  Taking into account
that ${H_f}^{-1} \sim |t_f| \sim \l_p$, and $L_f = L_i \,{a_f\over
  a_i} \sim |t_i| (|{t_i\over t_f}|)^{1/\sqrt3}$, one can write the
condition $L_f \ge e^{70} l_p$ in the following form:
\begin{equation}
\label{bound1}
 \left(\frac{\Phi_i}{\Phi_f}\right)^{1/\sqrt{3}} =
\left(\frac{|t_i|}{l_s}\right)^{1+1/\sqrt{3}} \ge e^{70} .
\end{equation}
It is important that these conditions should hold for any version of
the PBB universe, open, flat or closed. Moreover, one can show that
the same conditions follow from the requirement of flatness of the
universe, even if the homogeneity of the universe is ensured by some
other mechanism.  To be a successful model of inflation, the
pre-big-bang model must satisfy all of these inequalities together
with (\ref{exit}).

\section{Initial Conditions at the onset of inflation}\label{initial}

As we have mentioned in the previous section, the initial size of the
homogeneous inflationary domain $L_i$ should be greater than $t_i$.
Using Eq.~(\ref{bound1}), one finds the following lower bound on
$L_i$:
\begin{equation}
\label{size}
L_i \ge    e^{44}\, {l_s} \sim 2 \times 10^{19}\,   {l_s}.
\end{equation}
This is not much better than the situation in the non-inflationary big
bang cosmology, where it was necessary to assume that the initial size
of the homogeneous part of our universe was greater than $10^{30}
l_p$.

The measure of the curvature of the spacetime (the square root of the
square of the Riemann tensor) at this moment is given by ${\bf R} \sim
t_i^{-2} \lesssim 10^{-38} l_s^{-2}$.  Hence, in order to explain the
flatness of the universe in the PBB scenario one should start with a
universe which is flat from the very beginning.

To find the lower bound for the mass of the universe at the beginning
of PBB inflation, we consider the energy stored in the dilaton
field inside the horizon volume. Using equations (\ref{ee1}),
(\ref{sol1}), (\ref{exit}), (\ref{bound1}) and $L_i \sim |t_i|$, we
find
\begin{equation}
\label{mass}
M_i  \sim  \dot\sigma_{i}^2 L_i^3 l^{-2}_{p}
\sim  {|t_i|}{ l_p^{-2}}
 \ge  {e^{165}} M_s \sim {10^{72}}  M_s \ .
\end{equation}

One of the manifestations of the flatness problem in the usual big
bang theory is that the universe today is $10^{60}$ times heavier than
the natural mass scale $M_{p}$.  In order to resolve this problem in
the context of PBB cosmology, one should begin with a universe which
has the mass $M > 10^{72}\, M_s$, where $M_s$ is the only natural mass
scale at the PBB stage (and is approximately the same as the Planck
mass after the Big Bang, as we have discussed above).  Thus, PBB does
not seem to give us any real advantage in trying to explain the large
mass of the universe.

The best way to express the size of the universe in dimensionless
units is to calculate its entropy. For the universe filled with
ultrarelativistic matter the total entropy coincides, up to a factor
$O(1)$, with the total number of particles in the universe.  In
spacetimes possessing an event horizon, however, one has entropy even
in the absence of ordinary matter.  For example, the entropy of a
horizon-size domain of de Sitter space dominated by a scalar field
with a potential energy density $V(\phi)$ is proportional to the area
of the event horizon, $S = 2\pi A l_p^{-2} = 8\pi^2 l_p^{-2} H^{-2}$
\cite{GH}. This gives\footnote{To avoid confusion with the existing
  literature where this equation was derived, in this equation and in
  Eq.~(\ref{TUNN}) we use the ``gravitational'' Planck mass $M_p^2 =
  G^{-1} \sim 10^{19}$ GeV, rather than the ``stringy'' Planck mass
  $M_p^2 =(8\pi G)^{-1} $.  }
\begin{equation}\label{GH}
  S = {3 M_p^4\over 8 V(\phi)}\ .
\end{equation}

Similarly, we can estimate the entropy associated with the event
horizon in PBB inflation,
\begin{equation}\label{bh}
S = 2 \pi  {A_i}l^{-2} \ ,
\end{equation}
where $A_i$ is the area of the cosmological horizon at the beginning
of PBB inflation, and the parameter $l$ is the characteristic scale
for the problem.  We leave the option open that $l$ can be either
$l_s$ or $l_p$. When $l=l_p$, we see that since $A_i \sim 4 \pi L_i^2
\sim 4 \pi |t_i|^2 $, Eq.~(\ref{bh}) gives (using the conditions
(\ref{exit}) and (\ref{bound1}))
\begin{equation}
\label{bhval}
S  \sim  8 \pi^2  {|t_i|^2}{  l_p^{-2}}
\ge 8 \pi^2 e^{210}    \ge    10^{93}   \ .
\end{equation}
Hence the initial horizon entropy is exponentially large. In fact, it
is not very different from the total entropy of the observable part of
the present universe $S \gtrsim 10^{88}$.

One may argue, however, that perhaps in the stringy phase one should
calculate the area of the horizon in stringy units. If we take this
point of view, our estimate of the entropy becomes smaller, but still
remains extremely large:
\begin{equation}
\label{bhval2}
S  \sim 8 \pi^2  {|t_i|^2}{ l_s^{-2}}
\ge 8 \pi^2 e^{89}
\ge 2 \times 10^{38}   \ .
\end{equation}
Hence even this much weaker estimate gives an extremely large
quantity.

So far we have taken into account only the gravitational entropy.
Including any matter entropy would increase the total entropy. To
confirm this we can estimate the entropy of a universe which is
initially dominated by radiation with energy density $\rho_i$. The
total entropy $S_r$ within the horizon of initial size $|t_i|$ is
given by $S_r \sim (\rho_i t^4_i)^{3/4}$. From the constraint equation
in (\ref{ee1}) we find that $\rho_i t^4_i \sim t^2_i \Phi_i$. Using
the solution for radiation dominated PBB inflation obtained in
Ref.~\cite{tw} (case $b \gg 1$ in their notation), one can easily find
$t_i$ and $\Phi_1$, just as we did before for the case of
dilaton-dominated PBB inflation. This leads to the following
constraint on the total entropy of radiation in this scenario:
\begin{equation}
\label{radent}
S_r \gtrsim   10^{68}.
\end{equation}

The initial entropy can be used to evaluate the probability of
spontaneous formation of a homogeneous domain of an inflationary
universe. Indeed, one may argue that the entropy is the measure of
complexity of the universe, and the probability of creation of a
universe with a huge entropy should be exponentially suppressed,
\begin{equation}\label{TUNN}
P \sim e^{-S} = \exp \left(-{3 M_p^4\over 8 V(\phi)}\right) \ .
\end{equation}
This result exactly coincides with the result obtained in
Ref.~\cite{Tunn} by a different method. In inflationary cosmology it
implies that the probability of inflation starting in a state with
$V(\phi) \sim M_p^4$ is not exponentially suppressed. In other words,
one does not need to fine-tune initial conditions in the simplest
models of chaotic inflation, where inflation may start at $V(\phi)\sim
M_p^4$.

Meanwhile, the estimate $P \sim e^{-S}$ for the probability of quantum
creation of an inflationary universe leads to a vanishingly small
number in the context of the PBB scenario. For the dilaton dominated
regime we get, in the very best case,
\begin{equation}\label{TUNN2}
P \sim e^{-S} \lesssim \exp \left(-{10^{38}}\right) \ ,
\end{equation}
whereas for the radiation dominated universe the probability is even
much smaller,
\begin{equation}\label{TUNN2r}
P \sim e^{-S} \lesssim \exp \left(-{10^{68}}\right) \ .
\end{equation}

These arguments show just how serious the fine-tuning problem is in
PBB inflation. In order to be able to solve the horizon and flatness
problems, PBB inflation has to start from a universe which is already
very large, very dense and very homogeneous and isotropic. Its initial
entropy must also be exponentially large, suggesting that this initial
state is very improbable.

One can look at this issue from another point of view. Suppose we give
up the desire to solve all cosmological problems by a PBB stage, and
simply ask what is a natural duration of this stage. One can
immediately deduce the answer from Eq.~(\ref{bhval}). In order to
avoid exponential suppression of the probability of creation of a PBB
universe, it should be created at $|t_i |\lesssim l_p \sim M_p^{-1}$.
Thus, one may expect that the typical duration of PBB inflation is
given by the Planck time.

However, in our evaluation of the probability of formation of a
homogeneous inflationary domain of size $\sim 10^{19}$ $l_s$ we did
not take into account the possibility that such a state may naturally
appear as the result of some pre-inflationary dynamics. We will
investigate this possibility in the next two sections.

\section{Closed pre-big-bang universe}

In the usual big bang cosmology quantum fluctuations of the metric are
extremely large, the terms $\sim R^2$ in the effective action are
greater than $M^2_p R$, and the effective action approach breaks down
at the time $t \lesssim t_{ p} \sim M_{ p}^{-1}$.  Therefore one may
say that the universe (i.e., classical space-time) ``materializes'' at
the Planck time $t_p \sim M_p^{-1}$.  One could try to argue that in
the PBB theory the effective action approach breaks down near the big
bang, but not in the beginning of the PBB expansion. This seems to be
the case for the $(+)$-branch PBB solutions given in (\ref{sol1}).
They are described by the string effective action (\ref{act}) in the
far past, with ever weaker coupling and smaller curvature.  However,
if the universe is inhomogeneous, which is clearly a more generic
situation, the answer to this question can be completely different. We
should expect that, prior to inflation, the universe on a large scale
must be entirely inhomogeneous - after all, one of the roles of
inflation is to take one such universe and smooth it out by rapid
expansion. If such a solution had a curvature singularity in the past,
the action (\ref{act}) should be extended by higher derivative, or
higher order $\alpha'$, corrections, regardless of the magnitude of
the string coupling $g_s$. As a result, such a universe would be fully
shaped by string physics at the beginning, controlled by the scale
$l_s = \sqrt{\alpha'}$, and not by classical phenomena. The question
in this case is how big such a universe would have to be initially, in
order to survive until the beginning of inflation, and inflate to give
the post-big-bang.

The best way to investigate the naturalness of initial conditions in
inflationary cosmology is to study a closed universe, which at the
same time provides us with the understanding of the behavior of those
parts of the universe which are locally very dense. In the big bang
theory the total lifetime of a closed universe filled with radiation
is proportional to $M/M_p^2$, so a universe of the natural mass $\sim
M_p$ immediately collapses unless it begins inflating soon after its
formation. That is why the simplest models of inflationary theory
based on chaotic inflation are so much better than the new inflation
models from the point of view of initial conditions: In the chaotic
inflation models the process of exponential expansion may begin
immediately after the big bang \cite{book}.

Here we would like to analyze the same problem in the context of the
PBB theory. The exact spatially closed dilaton-metric solutions of
(\ref{ee1}) are \cite{tw}
\begin{eqnarray}
\label{curvsols}
&&a = \frac{\sqrt{B}l_p(0)}{3^{1/4}}\frac{(\cos
\eta)^{(1+\sqrt{3})/2}} {(-\sin\eta)^{(\sqrt{3}-1)/2}} \nonumber\\
&&~~~~~~~~\Phi = \frac{1}{l^2_p(0)} \Bigl(\frac{-\sin\eta}{\cos
\eta}
\Bigr)^{\sqrt{3}},
\end{eqnarray}
where $\eta$ is the conformal time, defined by $dt = a d\eta$. Here
$B=-a^3 \dot \Phi$ is a parameter which does not change during the PBB
evolution; $l_p(0)$ is the value of the effective Planck scale at the
onset of inflation.

In the limit $\eta \ll 1$, the solution (\ref{curvsols}) coincides
with the flat-space superinflating solution (\ref{sol1}), with $|t|
\sim |\eta|^{\sqrt{3}/(\sqrt{3}+1)}$.  The onset of inflation
corresponds roughly to $\eta \sim -\pi/4$, when $-\sin\eta \sim \cos
\eta $, and $a \sim -t_i \sim \sqrt{B} l_p(0)$, $\Phi_i \sim
l^{-2}_p(0)$. If we want the closed universe to be large enough to
incorporate a homogeneous inflationary domain of size $L_i \sim
2\times 10^{19}\ l_s$ (\ref{size}), one should have $ {B} \gtrsim 4
\times 10^{38}\ l^2_s/l^2_p(0) = 4
\times 10^{38}\ g_0^{-2} $. Taking into account that, according
to Eq.~(\ref{bound1}), $g_0^{-2} = {l^2_s/ l^2_p(0)} \sim {\Phi_i}/{\Phi_f} >
\exp({70\sqrt 3})$, one finds the following constraint on the
string coupling  constant $g^2 = l^2_p/l^2_s \sim {\Phi_f / \Phi_i}$ at the
onset of inflation:
\begin{equation}\label{g0}
g^2_0 < 10^{-53} \ .
\end{equation}
This leads to the following constraint on the parameter $B$:
\begin{equation}\label{BBBBBnew}
 B  \gtrsim  4\times 10^{38} g_0^{-2}  \ .
\end{equation}
For  the largest possible value $g^2_0 \sim 10^{-53} $ we get the constraint
\begin{equation}\label{BBBBB}
 B  \gtrsim  4\times 10^{91}  \ .
\end{equation}
For smaller $g_0$, the constraint on $B$ becomes even stronger.

The beginning of the evolution of the closed PBB universe is described
by Eq.~(\ref{curvsols}) in the limit $\eta \rightarrow - \pi/2$. In
this case the solution (\ref{curvsols}) can be approximated by the
solution (\ref{sol2}), where $t \sim (\pi/2 -
|\eta|)^{\sqrt{3}/(\sqrt{3} - 1)}$. Thus, $\eta \rightarrow - \pi/2$
again corresponds to $t \rightarrow 0$, but now the scale factor $a$
vanishes, and $R$ and $\Phi$ diverge (which implies that the string
coupling $g_s$ vanishes). Here $t$ corresponds to the time after the
{\it initial} singularity in the closed PBB universe. In what follows
we will concentrate on the investigation of conditions near this
singularity.

As we have discussed earlier, the description based on the effective
action (\ref{act}) breaks down when $R \sim l^{-2}_s \sim M^2_s$,
since the higher order $\alpha'$ corrections must be added regardless
of the value of the string coupling $g_s$. Further, it does not make
much sense to consider universes of size smaller than $l_s$ in the
effective action description, since in string theory we cannot fit any
low energy mode in such a universe. Hence, in this limit the solution
is valid until $a \sim l_s$ or $R \sim l^{-2}_s$, whichever comes
first when one goes toward the singularity at $t = 0$. From
Eq.~(\ref{sol2}) we see that $a \sim t^{1/\sqrt{3}}$ and $H^{-1} \sim
t$ at small $t$. Thus $H$ grows much faster than $a^{-1}$ in the limit
$t \to 0$. As a result, near the initial singularity the curvature $R
= 6 \dot H + 12 H^2 + 6k/a^2$ is dominated by the kinetic terms, $R
\sim H^2 \sim t^{-2} \gg a^{-2}$. This means that the effective action
description breaks down not at $a \sim l_s$, but at the stringy time
$t\sim t_s \sim l_s \sim M_s^{-1}$. Thus, the $k=1$ universe emerges
out of the initial stringy phase at the stringy time $M_s^{-1}$ after
the singularity.

We would like to point out that this is a general conclusion for all
stringy models where the universe experiences a power-law expansion $a
\sim t^\beta$ near the singularity, with $\beta = O(1)$. Indeed, in
this regime $H \sim t^{-1}$, and $R$ receives a contribution $ 12 H^2
\sim t^{-2}$, which becomes greater than $M_s^2$ for $t < M_s^{-1}$.
This means that the stringy time $t_s \sim M_s^{-1}$ plays the same
role in stringy cosmology as the Planck time in the standard big bang
theory.

If we take a $k=1$ PBB inflationary solution which satisfies all our
constraints (\ref{exit}), (\ref{bound1}), (\ref{size}), (\ref{mass})
and (\ref{bhval}), and extrapolate it backwards towards the initial
singularity, by the time the curvature again reaches the string scale
the universe will still be very large. Indeed, as we have already
mentioned, at $t \ll t_i \sim -\sqrt{B}l_p(0)$ one has $a \sim
t^{1/\sqrt{3}}$. Thus at $t \sim l_s$ one has
\begin{equation}\label{BEG}
a \sim  l_s\,
  (\sqrt{B}l_p(0)/l_s)^{1-1/\sqrt 3} \gtrsim 10^8\,  l_s \ .
\end{equation}

Note that if at the initial stage the size of the universe is $10^{8}$
times greater than the horizon $\sim t_s \sim l_s$, then one should
require that the universe is homogeneous in $10^{24}$ causally
disconnected domains of size $\sim l_s$. One can estimate the
probability of this event as
\begin{equation}\label{UUU}
P \sim \exp({-10^{24}}) \ .
\end{equation}
This clearly demonstrates that the condition (\ref{BEG}) requires the
universe to be unnaturally large from the very beginning.

If one wants to avoid this probability suppression, one needs to
consider a closed universe which consists of just one causally
connected domain of size $l_s$ at the stringy time $t_s \sim l_s$.
Such a universe would be immediately dilaton-dominated, would start
inflating - and would need to exit right away, after having been
inflating for an amount of time merely of order $l_s$. In this case
one would have $l_s \approx l_p$, so the total duration of the PBB
stage would be $ \sim M_p^{-1}$. Such a universe would never inflate
enough to solve the horizon and flatness problems.

Now let us estimate the total mass of the universe at this moment (at
the stringy time $t_s \sim M_s^{-1}$ after the initial singularity in
the PBB scenario). The calculation is essentially the same as the one
which leads to the estimate of the mass at the onset of inflation
(\ref{mass}). We again have
\begin{equation}
\label{mass2}
M_b \sim \dot \sigma_b^2 L_b^3 l^{-2}_p(b) \ ,
\end{equation}
where the index $b$ denotes quantities at the moment of birth of the
universe, at $t \sim t_s$.  A good estimate of the size of a closed
universe is $L_b \sim a$, where $a$ is given in Eq.~(\ref{BEG}). Using
$\dot \sigma_b^2 \sim R_b \sim l_s^{-2}$ and (\ref{BEG}), and
recalling that $B> 10^{91}$, we find that
\begin{equation}
\label{mass3}
M_b \sim \dot \sigma_b^2 L_b^3 l^{-2}_p(b)\sim M_s B
\gtrsim    10^{91} M_s \ .
\end{equation}
Thus the mass of the PBB universe at the moment of its creation is
even much greater than its mass at the moment when inflation begins
(\ref{mass}).

Similarly, one can obtain a constraint on the initial value of the
string coupling constant $g^2 = (\Phi l_s^2)^{-1}$ at the moment $t_s$
after the initial singularity:
\begin{equation}\label{g_b}
g_b^2 < 10^{-67} \ .
\end{equation}
This constraint may be viewed as very unnatural from the point of view
of M-theory.  Indeed, weak coupling limits of consistent string
theories are obtained as dimensional reductions of the $11D$ M-theory,
which admits $11D$ supergravity as its low energy limit. In the
process of dimensional reduction of M-theory to a string theory, the
string coupling (or equivalently the dilaton field) is identified with
the size of the eleventh direction. In particular, if ${\cal L}_{11}$
is the eleven-dimensional Planck length, we have the following
relationship between the size of the $11^{\rm th}$ direction and the
$10D$ dilaton field: ${\cal R}_{11} = {\cal L}_{11} \exp(\Sigma/3)$,
where ${\cal R}_{11}$ is the size of the $11^{\rm th}$ direction and
$\Sigma$ the $10D$ dilaton. In order to make contact with
four-dimensional physics, on which we focus on in this article, we
must further reduce the $10D$ string theory to a $4D$ one, which for
our purposes has been defined by the truncation of the action given in
(\ref{act}). This reduction in the context of the PBB cosmology is
usually performed on some internal rigid Calabi-Yau three-fold, and if
the volume of the Calabi-Yau three-fold is ${\cal V}_6$, we get the
following equation:
\begin{equation}
\label{ls}
l^2_s = \frac{{\cal L}_{11}^8}{{\cal V}_6} \ .
\end{equation}
Since the usual PBB scenario assumes that the Calabi-Yau is rigid
(i.e., has constant size), we can set $\sigma = \Sigma$.  The natural
volume of the Calabi-Yau three-fold is the string volume, ${\cal V}_6
\sim l_s^6$. We can see this if we recall that strings cannot probe
distances shorter than $l_s$, and hence it does not make sense to set
the volume of the Calabi-Yau to be less than $l^6_s$. Also, since
today the Planck scale is roughly equal to string scale, $l_p \sim
l_s$, if the volume were considerably larger than the string volume,
there would be Kaluza-Klein modes with masses much smaller than the
Planck mass today. To prevent this, for simplicity we can assume that
the Calabi-Yau manifold is compactified at the string scale.  In any
case, this will give us a rough order-of-magnitude estimate, and any
deviation from it could be absorbed in redefining $\sigma$.  A
possible subtlety involving $D$-branes and the fact they can probe
distances shorter than $l_s$ can be ignored since the PBB scenario is
defined in the perturbative sector of string theory, where $D$-branes
are super-heavy. All this leads to the conclusion that $l_s \sim {\cal
  L}_{11}$. Therefore, we find that when a successful PBB universe has
emerged from the initial string phase, it must have satisfied
\begin{eqnarray}
\label{r11}
 \frac{{\cal R}_{11}(b)}{a(b)} &\sim& \frac{l_s g_b^{2/3}}{a(b)} \sim
\frac{g_b^{2(\sqrt{3}+1)/3}}{g_0^{2/3}} \le e^{-10(\sqrt{3}+1)}
g_0^{2/\sqrt{3}} \nonumber\\ &&\le e^{-98} \sim 10^{-42} \ .
\end{eqnarray}
which clearly indicates that if viewed as an M-theory configuration,
the initial universe must be an extremely asymmetric one in order to
inflate.

\section{Open pre-big-bang universe}

The problems described in the previous section arise if the universe
is closed.  If the universe is initially flat or open, there is no
initial singularity at the stringy time $t_s \sim l_s$. Here we
briefly review the issue of naturalness of initial conditions for the
spatially open cosmologies.  Instead of carrying out a comprehensive
analysis, we merely outline generic kinematic conditions and indicate
the potential dangers.

The exact spatially open dilaton-metric solutions of (\ref{ee1}) are
\cite{tw}
\begin{eqnarray}
\label{curvsols1}
&&a = \frac{\sqrt{B}l_p(0)}{3^{1/4}}\frac{(\cosh
\eta)^{(1+\sqrt{3})/2}} {(-\sinh\eta)^{(\sqrt{3}-1)/2}}\ ,
 \nonumber\\
&&\Phi = \frac{e^{-\sigma}}{l^2_s} =
\frac{1}{l^2_p(0)} \Bigl(\frac{-\sinh\eta}{\cosh
\eta}
\Bigr)^{\sqrt{3}}\ .
\end{eqnarray}
The line element is
\begin{equation}
\label{metric}
ds^2 = a^2(\eta)\left(- d\eta^2 + \frac{dr^2}{1+r^2}
+ r^2 d\theta^2 + r^2
\sin^2  \theta \, d\varphi^2\right),
\end{equation}
where $\eta$ is the conformal time, defined by $dt = a d\eta$. As
earlier, we have $B=-a^3 \dot \Phi$; $l_p(0)$ is defined as
(approximately) the value of the effective Planck scale at the onset
of dilaton domination.

In the limit $\eta \ll 1$, the solution (\ref{curvsols1}) is again
approximated by the superinflating flat-space solution (\ref{sol1}),
with $|t| \sim |\eta|^{\sqrt{3}/(\sqrt{3}+1)}$.

The onset of PBB inflation occurs at the time $t_i \sim - \sqrt{B}
l_p(0)$.  This corresponds to $\sinh \eta \sim -1/\sqrt{\sqrt{3}+1}$.
At this time, the scalar field $\Phi$ is roughly $\Phi \sim 0.32\,
l_p^{-2}(0)$, and hence the string coupling is
\begin{equation}
\label{coupdd}
g =\frac{l_p(t)}{l_s} \sim 1.77 g_0.
\end{equation}

Let us now compare this to the conditions in the limit when the
universe is ``infinitely young", i.e., $\eta \rightarrow -\infty$. In
this limit, $-\tanh \eta \rightarrow 1$, and so $\Phi = 1/l_p^2(0)$,
giving for the initial string coupling $g_0 = l_p(0)/l_s \ll 1$.
Hence the string coupling $g$ is minimized by its initial value, in
contrast to what happens in the spatially closed and flat cases.

Since the inflationary stage of the open universe PBB theory does not
differ much from the inflationary stage of the closed universe
scenario, the constraints on the parameters $B$ and $g_0$ in the open
PBB universe remain practically the same as in the closed universe
case:
\begin{equation}\label{BBBB}
 B  \gtrsim  4\times 10^{38} g_0^{-2}   \ ,
\end{equation}
and
\begin{equation}\label{g0open}
g^2_0 < 10^{-53} \ ,
\end{equation}
so that
\begin{equation}\label{openphi}
\Phi_0 > g^{-2}_0  l_s^{-2} >10^{53} M_s^2  \ .
\end{equation}
For $g^2_0 = 10^{-53}$, the parameter $B$ should be greater than $10^{91}$.

In the limit $\eta \rightarrow - \infty$ we can approximate $-\sinh
\eta \sim \cosh \eta \sim e^{-\eta}/2$, which gives
\begin{equation}
\label{asyma}
a = \frac{\sqrt{B}l_p(0)}{2\cdot3^{1/4}} e^{-\eta} \ .
\end{equation}
Using $t = \int a(\eta) d\eta$, we find that up to an additive
constant,
\begin{equation}
\label{tim}
t = - \frac{\sqrt{B}l_p(0)}{2\cdot3^{1/4}} e^{-\eta} \ ,
\end{equation}
which yields $a = |t|$ and $H = - |t|^{-1}$ when $t \rightarrow
-\infty$.
{}From the relation $B=-a^3 \dot \Phi = -|t|^3 \dot \Phi$ it follows
that
\begin{equation}\label{INIT}
 \Phi(t)  =    \Phi_0 - {B\over 2 t^2} \ .
\end{equation}

Let us discuss the homogeneity problem in this scenario. Eq.~(\ref{INIT}) shows
also
that the difference between $\Phi_0$ and
$\Phi(t)$ in the early universe was very small:
\begin{equation}
\Delta\Phi =   \Phi_0 - \Phi(t)  = {B\over 2 t^2} \ .
\end{equation}
In the limit when $t \rightarrow -\infty$ (going backwards in time),
the spatial curvature $a^{-2}$ vanishes as $ t^{-2}$, whereas the
energy density is falling even faster: $\rho = {\dot\Phi^2 \over
  2\Phi} \sim {B^2\over 2\Phi_0} t^{-6}$. The same is true for the
Riemann tensor, $R_{\mu\nu\alpha\beta}$. It is completely determined by
the energy density, so it vanishes equally fast:
$R_{\mu\nu\alpha\beta} \sim {\dot\Phi^2 \over 2\Phi^2} \sim {B^2\over
  2\Phi_0^2} t^{-6}$.  Thus, the younger the universe, the flatter the
space-time.

As we have already mentioned, in order for the $k=-1$ PBB universe to solve the
homogeneity, flatness and horizon problems, at the onset of inflation the
linear dimension of a homogeneous and
isotropic patch which starts to inflate must be at least as big as
$L_0 \sim \sqrt{B} l_p(0)$. This happens at $t_i \sim -\sqrt{B}
l_p(0)$. In the past this domain should be much greater, of size $L(t)
\sim |t|$, because of the growth of the scale factor $a \sim |t|$.

Thus, the initial size of a homogeneous domain must be infinite.  One
may wonder whether it is a good idea to explain the large size of our
universe by assuming that it is infinite, and to explain its
homogeneity by assuming that the universe was homogeneous on an
infinitely large scale from the very beginning. The only consistent
version of an open universe theory which explains how a universe may
become homogeneous on an {\it infinitely} large scale is given by
inflationary cosmology \cite{Open}. In the models proposed in
\cite{Open}, the homogeneity of an open universe is explained in a
very nontrivial way by a preceding stage of indefinitely long false
vacuum inflation.  However, in our case we do not have any
pre-inflation before the pre-big-bang. Therefore we do not know why
the universe should be even approximately homogeneous from the very
beginning, so that ${\delta \rho\over \rho} \lesssim 1$ over an
infinitely large scale.

Indeed, as we have already mentioned, in the limit $t \to -\infty$ the
energy density of the matter and the curvature of the spacetime in an
open PBB universe were infinitesimally small. This means that the
universe was practically indistinguishable from empty Minkowski space.
It looks as an open universe only due to the presence of an
infinitesimally small amount of matter moving in a coherent way. Thus,
to produce a contracting open PBB universe one should take an empty
Minkowski space, add an infinitesimally small amount of nearly
homogeneously distributed dilaton field $\Phi$, and make this field
move in a coherent manner all over an infinite universe. This seems to
be much more complicated than to take a Planck-size domain filled with
a scalar field $\phi$ in a chaotic inflation scenario, and let this
domain inflate and self-reproduce, and create an infinitely large
amount of exponentially large homogeneous domains \cite{book}.

The possibility that the universe was homogeneous at $t \to -\infty$
may become even more complicated when one takes into account quantum
fluctuations of the field $\Phi$. Typically, one expects quantum
fluctuations to be important only near the cosmological singularity.
However, this is not the case in the open PBB universe. In the limit
$t \to -\infty$, the classical value of energy of the scalar field
$\Phi$ was vanishingly small, and quantum fluctuations could play a
dominant role.

Indeed, let us estimate a typical amplitude of quantum fluctuations on
a scale comparable with the initial size of our homogeneous domain
$L(t) \sim |t| \sim |H^{-1}|$. The equation of motion for the Fourier
modes of dilaton perturbations can be written as follows:
\begin{equation}
\label{perturbation}
\ddot \Phi_q + 3 H \dot \Phi_q + \frac{q^2}{a^2}\Phi_q =0 \ ,
\end{equation}
where $q$ is the comoving wavenumber, or equivalently, the inverse
comoving wavelength. At large $|t|$, the field $\Phi = l_p^{-2}$
approaches a constant limit $\Phi_0$, so we can ignore the variation
of the Planck mass.  Then, one can represent $\Phi$ as \ $l^
{-2}_{p}(1 + {\phi l_p})$, where $l_p \approx l_p(0)$ is approximately
constant. The action (\ref{act}) in terms of $\phi$ is $\int d^4 x
\sqrt{g}\, \{R\, l^{-2}_{p}/2 + (\nabla \phi)^2/2 + ...\}$. Thus the
field $\phi$ has canonical normalization. Hence its quantum
fluctuations on the scale $H^{-1}$ are roughly given by $\delta \phi
\sim {H\over 2\pi}$\, \cite{book}, and therefore we can write
\begin{equation}
\label{quantphi}
\delta \Phi \sim \frac{\delta \phi}{l_p} \sim
\frac{\sqrt{\Phi} H}{2 \pi} \sim \frac{1  }{2 \pi l_p(0)}
|t^{-1}| \  .
\end{equation}

These quantum fluctuations oscillate with a period $T \sim |t|$, so
they do not actually change much during the subsequent evolution of
the universe. Therefore they can hardly be distinguished from the
classical dilaton field $\Phi$. In order to appreciate the importance
of these quantum fluctuations, one should compare them with the
difference $\Delta \Phi = \Phi_0 - \Phi(t) = {B\over 2 }\,t^{-2}$.
This difference is the only feature which distinguishes the
contracting open PBB universe from Minkowski space. It is easy to
show that quantum fluctuations become much greater than $\Delta \Phi$
for
\begin{equation}\label{level}
|t| \gg B\, l_p(0) \ .
\end{equation}

One can reach a similar conclusion by comparing the energy density of
the homogeneous component of the scalar field, $\rho = {\dot\Phi^2
  \over 2\Phi} \sim {B^2\over 2\Phi_0} t^{-6}$, and quantum
fluctuations of the energy density of this field, which includes terms
such as ${\langle \dot\Phi^2\rangle\over 2\Phi} \sim {\langle
  (\delta\Phi)^2\rangle\over 2t^2 \Phi} \sim t^{-4}$. We see that the
energy density of quantum fluctuations on the scale $|t|$ is much
greater than the energy density of the homogeneous component of the
scalar field for $|t| \gg B\, l_p(0)$.

This means that at the very beginning of the evolution of the universe
in the PBB scenario, quantum fluctuations $\delta \Phi$ are always
much greater than $\Delta \Phi$, and their energy is much greater than
the energy of the field $\Phi$. They completely destroy the
homogeneity of the open universe PBB solution on the scale $|t|$ which
is important for the subsequent PBB inflation. In other words, the
assumption of initial homogeneity of the open PBB universe seems to be
internally inconsistent when quantum fluctuations are taken into
account.

We should emphasize that we are not discussing here the small density
perturbations ${\delta\rho \over \rho} \lesssim 1$ and small
perturbations of the metric in the open universe background, which can
be studied with the standard methods of perturbation theory.  We do not study
here the Jeans instability, which may or may not exist in the PBB cosmology.
Rather we argue that
the initial perturbations discussed above completely destroy the homogeneous
background, making it look like a collection of pieces of open
universes and closed universes with completely different density,
matched together to provide a chaotic distribution of segments of
space-time with different properties. It is correct that in the limit
$t \to -\infty$ the density of matter decreases, so we are rapidly
approaching Milne space, or Minkowski space \cite{buon}. The question,
however, is whether small perturbations of density in the limit $t \to
- \infty$ are greater than the density of the homogeneous component of
the field. If this is the case, then the simple description of the
universe in terms of Friedmann open or closed universe models becomes
invalid.  The way to see it is to compare the part of the Riemann
tensor induced by the homogeneous dilaton field,
$R_{\mu\nu\alpha\beta} \sim {B^2\over 2\Phi_0^2} t^{-6}$, with the
fluctuations of the Riemann tensor related to quantum perturbations,
which is proportional to $ \Phi_0^{-1} t^{-4}$.  Obviously, quantum
fluctuations completely change the geometry of space for $|t| \gg B\,
l_p(0)$.

Moreover, in order to ensure the homogeneity of an inflationary domain
of initial size $L_0 \sim \sqrt{B} l_p(0)$, one would need the
universe to be homogeneous not only on a scale $\sim |t|$, but also on
scales much greater than $|t|$.  Indeed, the conditions at any point
of the universe at a moment $|t_0|$ can be influenced by effects
occurring at a distance equal to its particle horizon,
\begin{equation}\label{hor}
L_h = a(t) \int_t^{t_0} {dt \over a(t)}
 \sim |t| \ln{|t|\over |t_0|} \gg |t| \
{}.
\end{equation}
As we have argued above, quantum fluctuations tend to destroy
homogeneity on a scale $\sim |t|$. Similar arguments show that quantum
fluctuations also destroy homogeneity on a larger scale $\sim |t|
\ln{|t|\over |t_0|}$.

In fact, perturbations which behave as $|t|^{-1}$ exist at the
classical level, too \cite{buon}, but their amplitude was unknown, so
it was hard to evaluate their significance.  One could do it only by
considering them as small perturbations on the open universe
background. As we believe, this is not the case for quantum
fluctuations, which can completely destroy this background.

Despite the chaos created by the large density perturbations discussed
above, one may expect that in an infinite universe there always are
many sufficiently large domains where the universe will be
sufficiently homogeneous \cite{buon}. In such domains inflation will
begin, just like in the chaotic inflation scenario \cite{book}, so one
may argue that one should concentrate on such domains and ignore those
domains which are too inhomogeneous to inflate.

Indeed, it is quite possible that despite initial inhomogeneities, the
universe always remains homogeneous on a stringy scale simply because
the scale of inhomogeneities in string theory can hardly become
smaller than $l_s$, just like the scale of inhomogeneity in gravity
theory cannot be smaller than the Planck scale.  However, as we have
discussed in section \ref{initial}, PBB inflation can solve the
cosmological problems only if the size of the initially homogeneous
domain is at least 19 orders of magnitude greater than the stringy
scale. At the moment we are unaware of any mechanism which would
ensure homogeneity of the universe on such a large scale.\footnote{In
  fact, {\it a priori} the scale $10^{19} \, l_s$ (unlike the scale
  $l_s$) does not look special in any respect. If there were any
  non-inflationary mechanism which would naturally produce homogeneous
  domains of size $10^{19} \, l_s$, then one would imagine that the
  same mechanism could possibly explain homogeneity of the universe at
  a much greater scale such as $\sim 10^{30} l_s$.  It is very hard to
  see how it could be otherwise - why would such a mechanism act
  selectively on inhomogeneities, smoothing ones at scales $\le
  10^{19} l_s$ and failing to smooth the ones that occur at scales
  $\ge 10^{30} l_s$. Then one would not need a prolonged stage of PBB
  inflation at all.}  Thus, we are returning exactly to the same
problem as in the case of the closed universe.

One could argue that in a certain sense this might not be such a great
problem.  The argument could go as follows: In the standard big bang
theory the universe was required to be homogeneous at the Planck time
on the scale $10^{30}\ l_p$. In the PBB cosmology the requirement is
much more modest; the size of a homogeneous domain should be greater
than $10^{19} \ l_s$. Thus, in the absence of
any natural model of
stringy inflation, this may still be considered a substantial
progress. Indeed, it seems much more probable to achieve homogeneity
in a domain of the
size $10^{19} \ l_s$ and then increase this size by the
PBB inflation up to $10^{30}\ l_p$, rather than to start from the very
beginning with a homogeneous domain of the
size $10^{30}\ l_p$.

However, this argument does not seem to help much. Indeed, if the
initial homogeneous domains were produced by chance rather than by
some additional stage of pre-inflation, then it seems much easier to
produce homogeneous domains of the
size $10^{18} \ l_s$ rather than the
homogeneous domains of the
size $10^{19} \ l_s$. Domains of the
initial size
$10^{18} \ l_s$ will inflate and produce locally homogeneous universes
with stars and planets like ours, but such universes will be extremely
inhomogeneous on the superlarge scale comparable to the scale of the
horizon $\sim 10^{28}$ cm.

This is the main reason why we need inflation, and why inflation must
somewhat ``overshoot.'' In the situation where a short stage of
inflation is more probable than a long stage of inflation, the
universe may be quite hospitable to the existence of life as we know
it, but it will be very inhomogeneous on the scale of the
horizon. This
would contradict observational data which show that the universe on
the scale of the
horizon is homogeneous at the level better than
$10^{-4}$.

It was mainly for this reason that all open inflationary universe
models proposed prior to \cite{Open} failed. It was always possible to
assume that inflation somewhat ``undershoots,'' and the universe
remains open if it was open from the very beginning. But this
``undershooting'' implied that inflation could not solve the
homogeneity problem. As we have already mentioned, in the models
proposed in Ref.~\cite{Open} this problem was solved by a prolonged
stage of inflation preceding the creation of a single-bubble open
universe. There is no such stage in the current version of the PBB
scenario.

%These constraints play a crucial role in our article,
%so we should perhaps
%emphasize again that they follow both from the requirement of homogeneity of
%the universe on a scale $10^{30}$ $M_p^{-1}$ at the Planck time, and,  {\it
%independently}, from the condition of flatness of the universe on this scale.
%Indeed, suppose that some unknown processes can make the universe entirely
%homogeneous at the beginning of the PBB inflation.

But what if, despite all our arguments, we were able somehow to solve the
homogeneity problem in the open PBB scenario?  Would it mean that we no
longer need to have the stage of PBB inflation at all, and our constraints on
$g_0$ and $B$, given by Eqs. ~(\ref{BBBB}) and (\ref{g0open}), disappear?

The answer is no. As we have
mentioned in Sect. \ref{initial}, we would still need
to have inflation of the same duration as
before.  If we are going to solve the flatness problem, we need the scale
factor of the open universe to become greater than $10^{30}$
$M_p^{-1}$ at the Planck time. Since this scale factor at the onset of
inflation is of the same order as  $|t_i| \sim   \sqrt{B} l_p(0)$ or greater,
the
constraints on $B$, $g^2_0$ and $\Phi_0$ which follow from the condition of
flatness of the universe coincide with the constraints, Eqs.~(\ref{BBBB}),
(\ref{g0open}), and
(\ref{openphi}), obtained from the condition of homogeneity of
the universe on a scale $|t_i|$.

Thus, even if one finds a way to ensure the homogeneity of
an open PBB universe on a scale much greater than $|t|$ without any
use of inflation, Eqs.~(\ref{BBBB}) and (\ref{openphi}), obtained from the
condition of flatness of the universe, will imply that the
solution $\Phi(t) = \Phi_0 - {B\over 2 t^2}$ describing initial
conditions in an
open universe must contain extremely large parameters
$\Phi_0 = g_0^{-2} M_s^2 > 10^{53} M_s^2$ and $B > 4\times 10^{38} g_0^{-2}
>  10^{91}$.

One might try to argue that the large values of these parameters
are somehow related to the postulated weakness of the
string coupling in the early
universe. There is a big difference, however, between a simple assumption that
$g^2$
was small in the PBB universe, and the requirement that it must be smaller than
$10^{-53}$. We are unaware of any   natural explanation of this number in the
PBB theory.
Moreover, even the extraordinary smallness of the string coupling   does not
help to solve the flatness problem. Indeed, if one keeps $B = const$, then in
the limit $g_0 \to 0$ the scale factor of the open universe at the beginning of
the PBB inflation vanishes, $a \sim \sqrt B g_0 \l_s \to 0$. Meanwhile we need
it to be greater than $10^{19}l_s$ to solve the flatness problem.  That is why
we get the
constraint $B  \gtrsim  4\times 10^{38} g_0^{-2} $, which implies that
even for $g_0 \sim 1$ we would need to have $B\gtrsim  4\times 10^{38}$, and
in the weak coupling limit $g_0 \to 0$ we would need to have an infinitely
large
$B$.

Obviously, the requirement that $g^2_0$ should be smaller than $10^{-53}$
and $B$ should be greater than $  4\times 10^{38} g_0^{-2}$ is not a solution
of the flatness
problem, but its reformulation, where instead of one problem we have two.

After having investigated closed and open PBB universes, we will
briefly discuss the spatially flat universe. Such a universe can be
viewed as a special limit of the $k=-1$ cases where the instant of
dilaton domination has been pushed off to $-\infty$.  This suggests
that the $k=0$ cases are very special to start with. To qualify this
statement, one can view the flat cases as asymptotic repellers in the
phase space of the PBB theory. Indeed, if we take an arbitrary PBB
universe at a time well into the superinflating phase, and perform the
time-reversal transformation, we will note that the universe is
undergoing decelerated contraction towards the ``future".  This period
of decelerated contraction will last forever only if the ``initial
conditions" are infinitely fine-tuned (corresponding to the $k=0$
cases). Indeed, the curvature tensor of a flat universe vanishes at
$t\to -\infty$, so the flat universe expansion will be destroyed by
any finite perturbation at $t \to -\infty$, just like in the open
universe case. In a generic case, different parts of the universe will
either bounce into a phase of decelerated expansion ($k=-1$) or will
speed up into a phase of accelerated contraction ($k=1$).

It is interesting to look at the flat universe regime from the point of view
of the behavior of the string coupling. In a closed universe the initial value
of $g$ is infinitely large. In an open universe it is small and finite. The
flat
universe case can be obtained from the open universe in the limit  $g_0  \to 0$
and $a_0  \sim \sqrt B g_0  l_s \to \infty$, i.e. $B g_0^2 \to \infty$. In
other words,  $B$ must be double-infinite. The fine-tuned nature of this limit
is evident.

A flat universe looks especially peculiar from the point of view of
M-theory.  Using the same arguments as in the preceding section, we
can easily show that the size of the $11^{\rm th}$ dimension is ${\cal
  R}_{11} \sim l_s g^{2/3}$, where g is the string coupling. In a flat
universe $g \to 0$ in the limit $t \rightarrow -\infty$.  This shows
that from the point of view of M-theory the whole universe should have
evolved from a single infinitely large and homogeneous $10D$ plane of
vanishing thickness. This is an infinitely anisotropic state from the
point of view of $11D$ geometry, which appears to be even less natural
than the $k=1$ case.

\section{PBB cannot self-reproduce}

One of the most interesting features of many versions of inflationary
cosmology is the possibility of eternal inflation due to the
self-reproduction of inflationary domains. The most dramatic
realization of this possibility appears in chaotic inflation
\cite{al}, but the self-reproduction of the universe is possible in
the new inflationary scenario as well \cite{Vilen}.

If PBB could self-reproduce, it could spontaneously start by a quantum
fluctuation in some region of the universe, no matter how unlikely
this is. Then, this region could give birth to other PBB-inflating
domains, thus eventually filling a larger and larger portion of the
whole universe with inflating regions, and circumventing the
naturalness problem we discussed in the previous section. Here we will
show that this does not happen.

The process of self-reproduction of inflation is based on
quantum-mechanical violations of energy conservation. In models such
as chaotic inflation, quantum fluctuations can (over)compensate the
decrease of the effective ``vacuum" energy due to the slow roll of the
order parameter, by pushing the order parameter back towards the
region of large effective mass faster than it rolls down the potential
ridge. In PBB, however, the energy which drives the expansion of the
universe is provided by the kinetic energy of the rolling string
coupling, and the duration of inflation is limited by requiring that
the coupling must be at most of order unity. Hence, the
self-regeneration of PBB would require that quantum fluctuations can
decrease string coupling $g_s$, or equivalently, increase the
effective Planck mass squared $\Phi = l_s^{-2} g_s^{-2}$ faster than
it rolls down during the expansion.

To give a qualitative description of the conditions for
self-regeneration, we should look at the dynamics of coupling
inhomogeneities. As we have indicated earlier, during the
dilaton-dominated epoch, the coupling obeys a simple Klein-Gordon
differential equation: $\nabla^2 \Phi = 0$.  The individual modes
$\Phi_{q}(t) \exp(i \vec q \cdot \vec x)$ propagate according to
Eq.~(\ref{perturbation}). The physical wavelength $\lambda$ of the mode
$\Phi_q$ and the comoving wavelength $1/q$ are related by $\lambda =
a/q$. Now, in PBB, $H \sim 1/|t|$ and $1/a^2 \sim |t|^{2/\sqrt{3}}$.
Thus, all waves with the comoving wavelength $1/q \ge
(l_s/|t_0|)^{1+1/\sqrt{3}}$ will exit the Hubble volume of the
universe at some time. Recall that the time and the background value
of the scalar field $\Phi$ are related by (\ref{mc+}). The metric and
the Hubble parameter are uniquely determined by $\Phi$, as can be seen
from (\ref{sol1}) and (\ref{mc+}). We can therefore parameterize the
instant when the wave exits by the magnitude of the field $\Phi$. For
a wave with the comoving wavelength $1/q$ this happens when the field
$\Phi(q)$ is such that the wave vector and the physical wavelength are
\begin{eqnarray}
\label{wavevecs}
q &=& a(\Phi(q)) H(\Phi(q)) \ , \nonumber \\
\lambda &=& \frac{a(\Phi(q))}{q} = \frac{1}{H(\Phi(q))} \ .
\end{eqnarray}
After this moment, as we see from equation (\ref{perturbation}), these
modes freeze out: since the physical frequency $q^2/a^2$ is subleading
to the damping term $\sim H$, inside the Hubble volume the waves with
wavelength greater than $1/H$ do not oscillate any more. Now, the
self-regeneration of inflation requires that at some later instant,
when $\Phi(\bar t) = \bar \Phi$, there must be a region inside the
original domain which looks exactly the same as the original domain
did when $\Phi(t)=\Phi$. Modeling the self-regeneration processes by a
simple doubling of domains, we see that $\bar \Phi$ is related to
$\Phi$ as the instant when the wave with the comoving wavelength
$1/\bar q = 1/(2q)$ froze out. Using this and the definitions
(\ref{wavevecs}), we see that $\bar \Phi$ is determined by
\begin{equation}
\label{tdouble}
a(\bar \Phi) H(\bar \Phi) = 2 a(\Phi) H(\Phi) \ ,
\end{equation}
or, explicitly, using (\ref{sol1}) and (\ref{mc+}),
\begin{equation}
\label{tdoubleexp}
\bar \Phi = \frac{\Phi}{2^{\sqrt{3}}} \ .
\end{equation}

By this time, the square of the effective Planck mass decreases by an
amount
\begin{equation}
\label{delpl}
\Delta \Phi = \Phi - \bar \Phi = \Phi
\Bigl(1 - \frac{1}{2^{\sqrt{3}}} \Bigr) \ .
\end{equation}
This corresponds to the classical rollover and must be compensated by
quantum fluctuations in order for the self-reproduction to proceed.
The amplitude of quantum fluctuations was estimated in the previous
section:
\begin{equation}
\label{quantphi1}
\delta \Phi \sim \frac{\delta \phi}{ l_p} \sim
\frac{\sqrt{\Phi} H}{2 \pi}\  .
\end{equation}
If the magnitude of the quantum fluctuation $\delta \Phi$ is greater
than the magnitude of the rollover $\Delta \Phi$ for any background
value of $\Phi$ and $H$, the self-regeneration can occur. However,
using (\ref{delpl}) and (\ref{quantphi1}), we find
\begin{equation}
\label{inphi}
{\delta \Phi\over \Delta\Phi} \sim
  \frac{2^{\sqrt{3}-1}}{ \pi(2^{\sqrt{3}}-1)}
\frac{H}{\sqrt{\Phi}}
 \sim   \frac{2^{\sqrt{3}-1}}{\sqrt{3}\pi(2^{\sqrt{3}}-1)}
 \frac{l_{p}}{|t|}\
{}.
\end{equation}
This ratio remains exponentially small until the very end of the PBB
stage, i.e.,
\begin{equation}
\label{phiineq}
\delta \Phi \ll \Delta \Phi \ .
\end{equation}
We see that quantum fluctuations at the stage of PBB inflation are
never large enough to overtake the rolling of the field $\Phi$.
Therefore eternal inflation is impossible in the PBB scenario, so it
cannot alleviate the problem of initial conditions in the PBB
cosmology.

\section{Conclusions}

In this article, we have presented arguments clarifying the problem of
initial conditions in the PBB scenario of inflation. Our results
suggest that the current version of the PBB scenario, described by an
effective action with higher order corrections within a single string
theory, cannot solve the homogeneity, isotropy, flatness and horizon
problems in the way these are solved by the usual inflationary
scenario.  The PBB universe must be huge and homogeneous from the very
beginning.  Our calculations show that in order to solve the horizon,
flatness and homogeneity problems, the initial size of the
inflationary domain in the PBB scenario should be at least 19 orders
of magnitude greater than the string length $l_s$, which is the only
natural length scale in the theory.  If the universe is closed, and is
to inflate enough to solve the horizon and flatness problems at the
Planck time, its initial size prior to PBB inflation has to be at
least $10^{8}$ times greater than the horizon size at the time of the
universe creation. The initial mass of such a universe has to be
$10^{91}$ times greater than the string mass. We have estimated the
probability of one such event to be $P \sim \exp(-10^{24})$, i.e.,
exponentially small. If one wants to avoid this strong suppression,
then the natural duration of PBB inflation appears to be as short as
$M_p^{-1}$.

If the universe is open, it must begin in a state with a vanishingly
small density of the dilaton field, which should be nearly
homogeneously distributed over an infinitely large length scale. We do
not know whether the assumption of initial homogeneity in an
infinitely large volume is a good way to explain the homogeneity of
our part of the universe.  Also, the initial homogeneity can be
destroyed by any finite density perturbations. Moreover, we argued
that the open universe solution is unstable with respect to quantum
fluctuations which can make spacetime completely inhomogeneous, so its
description in terms of an open universe may become inadequate.  Parts
of the universe similar to the one we inhabit can appear as a result
of PBB inflation in homogeneous domains with an initial size greater
than $10^{19} \ l_s$. Such domains may appear by chance in an infinite
inhomogeneous universe. It seems much easier, however, to produce
homogeneous domains of a smaller size. This means that typical domains
produced by PBB inflation may be homogeneous on a small scale, but
should be very inhomogeneous on the scale comparable to the size of
the observable part of our universe.

Even if it were possible to solve somehow  the problem of initial
homogeneity of an open PBB universe, there is one more problem. The
open universe solution is characterized by two dimensionless parameters,  $g_0$
and $B$.
If the flatness problem is to be
solved by PBB inflation, i.e. if we want to explain why the present value of
$\Omega$ is not vanishingly small, one should have $g_0^{-2} \gtrsim 10^{53}$
and $B
\gtrsim 10^{38}g_0^{-2} > 10^{91}$.
Note that to solve the flatness problem one should explain why the scale factor
of the universe at the Planck time was $10^{30}$ times greater than the Planck
length. Now in order to explain the origin of the large number $10^{30}$ in the
context of the PBB cosmology we must introduce two other numbers, $g_0^{-2}$ and
$B$, which should be greater than $10^{53}$ and $10^{91}$ respectively. We do
not see any natural explanation for
appearance of these two different large dimensionless parameters in the theory.

The problems mentioned above are further exacerbated by the fact that a regime
of eternal inflation is impossible in the PBB scenario. Hence, even if a
region of the universe began to inflate in a PBB phase, it would
remain solitary and isolated, and hence just as unlikely to produce
the universe we live in.

In closing, we note that the origin of the fine-tuning seems to be
twofold. First, PBB is {\it borderline} inflation, which does not
satisfy the standard inflationary condition $\dot H \ll H^2$. Second,
the scenario is defined to be entirely stringy, i.e., the coupling is
at most of order unity and so the full evolution of the universe from
its birth through PBB to the present day is entirely within the phase
space of a single string theory. The first property is a generic
feature of any superinflationary model, and thus cannot be atoned. In
contrast, one could attempt to relax the second condition and allow
the coupling to get greater than unity by the time the exit occurs.
While this might have sounded heretical prior to the development of
the ``web of dualities" \cite{w}, one could imagine that this region
of very large coupling $g_s \gg 1$, should be described by a weak
coupling region of an $S$-dual of the original string theory in which
PBB has begun, or alternatively, by a phase of the parent M-theory.
Even if this were permitted, however, the original coupling would have
to grow to about $10^{46}$ in order to remove the naturalness bounds
we have found. This can be seen, for example, from the mass bound
(\ref{mass}), which can be rewritten as $M\ge 10^{91} M_s g_f^{-2}$ where
$g_f$ is the coupling of the original theory at the time of the exit.
Moreover, even if this were allowed, the current initial conditions
require that the cosmological evolution must begin when at least one
of the compact dimensions is about $10^{42}$ times smaller than the
length scale of the space-time.

We must admit that independently of the prospects of the current
version of the pre-big-bang scenario, its investigation has shown us
several intriguing possibilities.  The dynamics of the PBB model with
an account taken of initial inhomogeneities produced by quantum
fluctuations deserves a more detailed investigation.  There is always
a chance that we could have missed something important, or that our
understanding of inhomogeneous PBB models is incomplete. One should
remember that in the very beginning of the development of inflationary
cosmology some of its authors claimed that it does not work and cannot
be improved \cite{GW}. It is not inconceivable that some properties of
the PBB models which look unnatural from our point of view can become
natural when seen from another perspective.  It would be very
desirable to find a way to overcome the problems which we have found,
because it could be an important step towards a realization of
inflationary cosmology in the context of string theory.

\bigskip
\section{Acknowledgements}
The authors are grateful to Ram Brustein and Gabriele Veneziano for
stimulating discussions and useful comments. This work was
supported by NSF grant PHY-9219345.

\end{document}